# Planning for climate neutrality in the Nordic power sector: Insights from a non-harmonised comparison of eight energy system models

**Authors:**

Emir Fejzic [a,*], Will Usher [a,b], Ida Græsted Jensen [c], Marianne Zeyringer [d], Oskar Vågerö [d], Maximilian Roithner [d], Guillermo Valenzuela-Venegas [d], Rasmus Bramstoft [e], Marie Münster [e], Jean-Nicolas Louis [f], Pernille Seljom [g], Miguel Chang [g], Eirik Ogner Jåstad [h], Dmitrii Bogdanov [i], Christian Breyer [i]

[a] KTH, School of Industrial Engineering and Management (ITM), Energy Technology, Energy Systems, Brinellvägen 68 SE-100 44 Stockholm, Sweden
[b] Open Energy Transition, Königsallee 52, 95448 Bayreuth, Germany
[c] Energy Modelling Lab, København K, Denmark
[d] Department of Technology Systems, University of Oslo, Kjeller, Norway
[e] Department of Technology, Management and Economics, Technical University of Denmark, Produktionstorvet, Bygning 424, Kongens Lyngby, Denmark
[f] Energy Research Area, VTT Technical Research Centre of Finland Ltd, Tekniikantie 21, 02044 Espoo, Finland
[g] Department of Energy Systems Analysis, Institute for Energy Technology (IFE), Post Box 40, 2027 Kjeller, Norway
[h] Faculty of Environmental Science and Natural Resource Management, Norwegian University of Life Sciences (NMBU), Norway
[i] LUT University, Yliopistonkatu 34, Lappeenranta, Finland
* Corresponding author, E-mail address: fejzic@kth.se (E. Fejzic).

## Email addresses of all authors

Emir Fejzic (fejzic@kth.se); Will Usher (will.usher@openenergytransition.org); Ida Græsted Jensen (Ida@energymodellinglab.com); Marianne Zeyringer (marianne.zeyringer@its.uio.no); Oskar Vågerö (oskar.vagero@its.uio.no); Maximilian Roithner (maximilian.roithner@its.uio.no); Guillermo Valenzuela-Venegas (g.a.v.venegas@its.uio.no); Rasmus Bramstoft (rabpe@dtu.dk); Marie Münster (maem@dtu.dk); Jean-Nicolas Louis (Jean-Nicolas.Louis@vtt.fi); Pernille Seljom (Pernille.Seljom@ife.no); Miguel Chang (Miguel.Chang@ife.no); Eirik Ogner Jåstad (eirik.jastad@nmbu.no); Dmitrii Bogdanov (dmitrii.bogdanov@lut.fi); Christian Breyer (Christian.Breyer@lut.fi)

## ORCID

Emir Fejzic (0000-0002-2489-8455), Will Usher (0000-0001-9367-1791), Ida Græsted Jensen (0000-0001-7409-3439), Marianne Zeyringer (0000-0003-1756-1878), Oskar Vågerö (0000-0002-0806-0329), Maximilian Roithner (0009-0008-5232-1996), Guillermo Valenzuela-Venegas (0000-0003-2067-7257), Rasmus Bramstoft (0000-0002-8875-8718), Marie Münster (0000-0002-5543-547X), Jean-Nicolas Louis (0000-0002-7917-9634), Pernille Seljom (0000-0002-3124-1056), Miguel Chang (0000-0001-7673-0622), Eirik Ogner Jåstad (0000-0002-1089-0284), Dmitrii Bogdanov (0000-0001-7136-4803), Christian Breyer (0000-0002-7380-1816)

## Keywords

Energy system modelling, Power sector transitions, Multi-model comparison, Climate neutrality, Nordic countries

## Highlights

- Eight non-harmonised models compare Nordic power-sector pathways
- Outputs are aligned to a common reporting basis across four countries
- Wind power is the clearest shared feature of 2050 transition pathways
- CCS, nuclear, and emissions outcomes vary strongly across models
- Transparent scope and assumptions are key for policy interpretation

## Abstract

The Nordic countries have adopted ambitious climate targets that require far-reaching power-sector transformations, making energy system modelling an important input to long-term planning. However, model-based evidence is produced using different model structures, assumptions, scopes, and scenario designs. This paper examines what can be learned from comparing such independently developed scenarios by assessing Nordic power-sector climate-neutrality pathways across eight structurally diverse energy system models. The comparison covers Denmark, Finland, Norway, and Sweden for 2030, 2040, and 2050, and focuses on electricity demand, generation capacity, CCS deployment, and power-sector $CO_2$ emissions. Inputs are not harmonised; instead, outputs are compared using a common reporting basis reflecting how modelling evidence is encountered in applied policy contexts. The results show broad agreement on the direction of transition. Wind power, mainly onshore but complemented by offshore wind in some countries, is the clearest cross-model finding and forms the backbone of the Nordic power system by 2050. At the same time, installed capacities, CCS deployment, nuclear outcomes, and emissions levels vary substantially. These differences are interpreted considering renewable-resource potentials, technology availability, policy constraints, sectoral and geographical scope, emissions-accounting boundaries, and different implementations of climate-neutrality targets. The study shows that non-harmonised model comparisons can support policy analysis by identifying where models point in the same direction, such as wind expansion, and where outcomes depend more strongly on model and scenario assumptions, such as CCS, nuclear, and net-negative emissions. For policy use, the findings underline the need to report model scope, technology representation, policy constraints, and emissions-accounting boundaries transparently.

# 1. Introduction

The Nordic countries of Denmark, Finland, Iceland, Norway, and Sweden have ambitious climate targets that require transforming their energy sectors [1]. With the aim of net-zero greenhouse gas (GHG) emissions by 2045, for example, Denmark [2] and Sweden [3] have objectives that surpass those envisaged in the European Union's (EU) Fit for 55 package [4], and Finland aims for carbon neutrality by 2035 [5], with Norway being an exception as it aims to reach a GHG emissions reduction of 90-95% by 2050 but has not set a year to reach net zero emissions. Such Nordic efforts are facilitated by tangible policy measures, viz., carbon taxes. In 1990, Finland became the first country in the world to levy a carbon tax, followed by Sweden (1991), Norway (1991), and Denmark (1992) [6]. As of 2025, the Nordic countries have some of the highest carbon taxes in Europe, with Sweden topping the chart at 134 EUR per tonne of $CO_{2eq}$ [7]. In addition to policy support, the Nordic power sector, regarded as the backbone of the regional energy transition, benefits from its vast hydro reservoirs and strong regional market integration [8]. In 2023, Norway generated 89% of its electricity from hydropower, Sweden relied on a diversified mix, Denmark mostly on wind power, and Finland on a balanced nuclear–hydro–wind–biofuel portfolio [9], as shown in Figure 1.

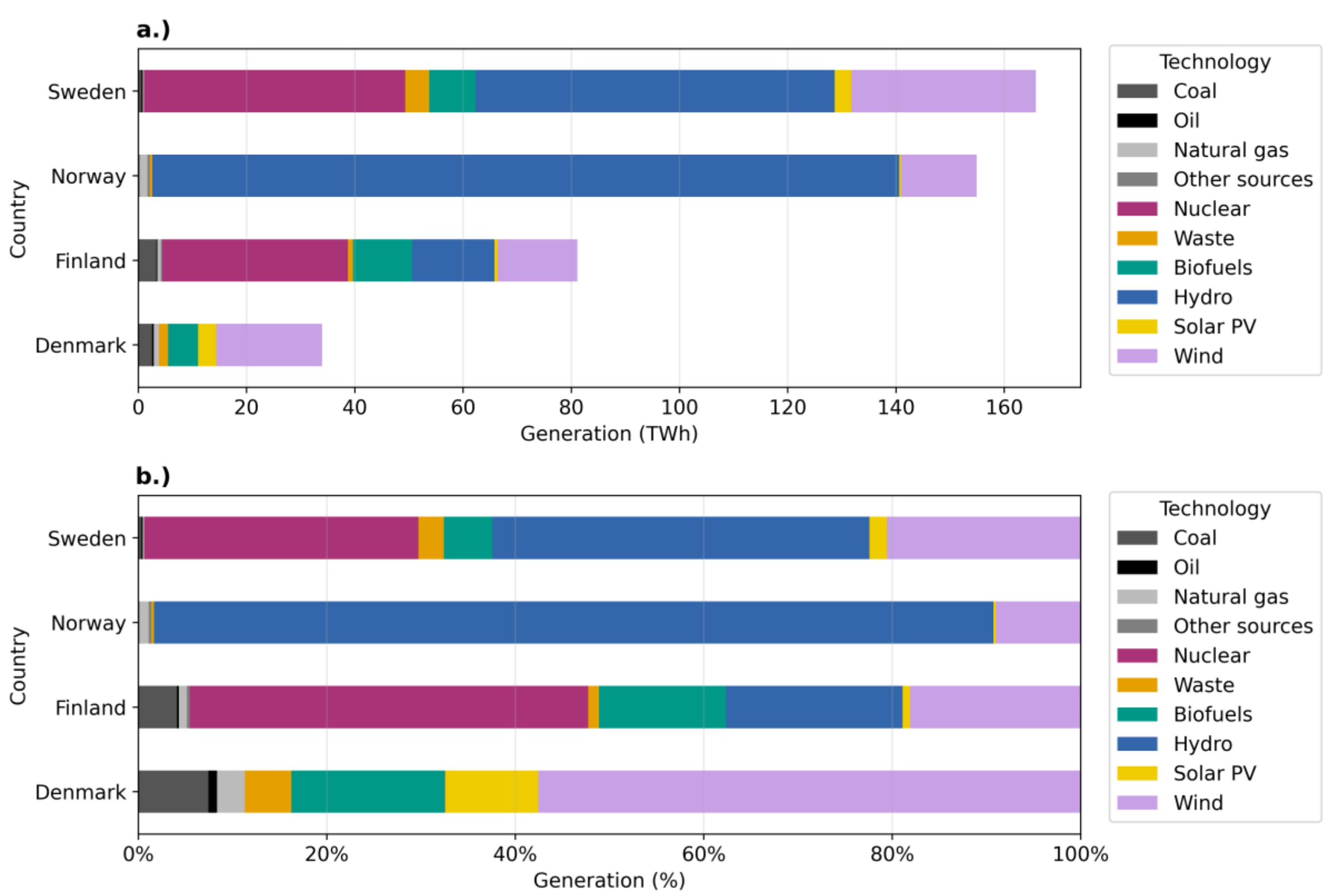

*Figure 1 Electricity generation of Nordic countries in 2023 (data from IEA used under CC-BY 4.0 license). Panel a.) shows the total electricity generation by technology expressed in TWh, while panel b.) illustrates the corresponding technological shares of national electricity generation.*

Energy system modelling is central to policy design in the Nordic region, yet the diversity of modelling approaches, assumptions, and data complicates comparability and reproducibility. Policymakers rely on model-based evidence to inform long-term strategies, while modelling choices are shaped by policy priorities and expectations [10, 11]. Credible policy measures require reproducible and coherent results across different models to instil confidence among stakeholders who will be affected by the sustainable energy transition [10]. This interdependence highlights the importance of transparent assumptions and explicit communication of model uncertainties and limitations, especially since model complexity can reduce transparency, and policymakers frequently influence data and assumptions used in modelling [11]. Existing model comparisons at the European and the Nordic scale can show the diversity of options for decarbonising the energy system when considering similar overarching narratives [12], and often rely on harmonised scenarios [13, 14] and input data [15, 16], or use coarse regional aggregations that obscure national differences [17].

Integrated assessment models have long been used to explore global climate-mitigation pathways [18] and have an established tradition of model intercomparison exercises [19], including dedicated efforts such as Energy Modeling Forum 14 [20]. However, despite their use in international policy assessment processes [21] and IPCC scenario assessments [22], their coarse regional aggregation and limited representation of emerging technologies constrain their usefulness for national and sub-national planning [23]. In contrast, detailed energy system models (ESMs) provide the technological, temporal, and geographical resolution needed for country-level decarbonisation analysis [24], but differences in model structure, assumptions, and input data can still limit comparability across results, motivating the multi-model approach adopted in this study. The Nordic region also has a long tradition of high-resolution modelling of low-carbon futures. The first conceptualisation of a carbon-neutral Danish energy system was proposed as early as 1975 [25], and interest in 100% renewable energy systems has since grown into a mature research field [26].

The integration of energy sectors is seen as an essential enabler for deep defossilisation and 100% renewable energy systems [27-29]. European climate-neutrality scenarios consider this as a key source of system flexibility and resilience [30-35], and high penetration of variable renewable energy (VRE) is a common feature across national energy system studies in the Nordics [36]. In Nordic countries, cross-sectoral integration is already an exploited feature that can be expected to play an even bigger role in the future [37], it provides flexibility to the system [38] and is often included across the different low-carbon scenarios portraying the Nordic energy system's transition. A recent overview of 100% renewable energy system studies for all Nordic countries [36] identified 69 studies by mid-2021, with additional studies in which the Nordic countries are part of more comprehensive analyses of the EU or Europe as a whole. Multiple low-carbon scenarios exist for both the Nordic region [39, 40], and its individual countries [41-43], yet these studies differ substantially in scope, assumptions, and methodological framing. A recent overview on highly renewable energy system analyses for the Nordic countries identified more than 150 studies, with Denmark the by far most studied country and EnergyPLAN and Balmorel the most used energy system modelling tools [44].

Comparable multi-model exercises have been undertaken in North America [45, 46], the European power market [47], and national analyses for Germany [48], the United States [49], Switzerland [50], and Japan [51], demonstrating the value of comparing structurally diverse modelling frameworks. In the Nordic region, coordinated initiatives such as the Nordic Clean

Energy Scenarios (NCES) programme [52], and national multi-model efforts linking energy system models with sector-specific models—for example, TIMES-Norway coupled with detailed transport models for road freight in Norway [53] provide important complementary insights. However, these approaches generally combine or align models rather than systematically compare independent energy system models under differing native assumptions. To date, only one study has performed a dedicated Nordic power sector model comparison [54], which focused on a narrow set of harmonised export and defossilisation scenarios. Consequently, despite substantial Nordic research on modelling future low-carbon energy system scenarios, a systematic country-level comparison across diverse frameworks remains absent.

Despite extensive modelling efforts, several gaps remain: (1) the absence of a country-level, multi-model assessment quantifying how Nordic energy system scenarios differ across diverse modelling frameworks when inputs are not harmonised; (2) limited systematic analysis of the extent to which modelled outcomes converge or diverge, and the structural or assumption-driven factors underlying these differences; and (3) transparent, reproducible comparisons that clarify how structural features of different modelling frameworks influence policy-relevant insights. Addressing these gaps is essential for improving confidence in energy system modelling as a basis for Nordic policy design.

Harmonised multi-model comparisons are widely used to improve comparability across models by aligning key inputs, assumptions, and scenario definitions [17]. However, such coordination does not eliminate all variation across models. Differences in model structure, scope, and formulation can continue to shape outcomes even in coordinated comparison exercises [14]. At the same time, energy system models are closely linked to policy processes, and their interpretation depends not only on methodological consistency but also on transparency regarding underlying assumptions [11]. In this context, comparing independently developed scenarios can provide insight into the range of outcomes that arise when different modelling approaches, scopes, and input data are applied to similar policy questions. The present study adopts this perspective by analysing non-harmonised model results, not as an alternative to harmonised intercomparison, but as a complementary approach that reflects how model-based evidence is typically encountered in applied policy settings.

In response, we conduct a multi-model comparison of the Nordic transition to climate neutrality, comparing eight structurally diverse models, Backbone Northern European model (Backbone-NEM) [55, 56], Balmorel-DTU/Balmorel-NMBU [57], ON-TIMES [58], IFE-TIMES-Norway [59], OSeMBE [60], highRES [61], and LUT-ESTM [62], within one overarching 2050 net-zero framing. By retaining native assumptions rather than imposing extensive harmonisation of inputs, the ensemble captures both structural and assumption-driven effects. This reflects how modelling results are produced and interpreted in practice and allows the analysis to focus on how differences between models influence the range of outcomes. We also answer broader calls for systematic inter-model comparison and transparent reporting [63]. To promote transparent reporting, all outputs are publicly available on Zenodo.

This study addresses the following research question: How do modelled Nordic power sector transitions differ across eight structurally diverse frameworks operating without harmonised inputs, and what explains convergence or divergence in outcomes? We compare key indicators, trace results to model structures and assumptions, and discuss implications for transparent and interpretable modelling for policy use. By directly linking the research question to the identified gaps, this study aims to clarify both the extent of convergence across models and the underlying sources of divergence, while maintaining consistency with real-world modelling practice.

# 2. Methods

This section describes the modelling frameworks included in the analysis and summarises their main characteristics. The focus is on key aspects such as model scope, structure, temporal resolution, and the implementation of net-zero constraints. Together, these elements provide the basis for interpreting similarities and differences across model results.

## 2.1. Models included

The models were selected because they provide recent climate-neutrality or near-net-zero scenarios with extractable power-sector results for at least one Nordic4 country and are actively used in Nordic or European energy-system analysis. The ensemble is intended to capture methodological diversity rather than statistical representativeness, spanning electricity-sector and full-energy-system models, national to European geographical scopes, hourly and time-slice temporal representations, and different net-zero implementations. The study therefore does not benchmark a complete population of Nordic models under identical assumptions but compares independently developed model scenarios on a common reporting basis to examine where Nordic power-sector transition insights converge or diverge. A summary of model characteristics is provided in Supplementary Table S1.

### 2.1.1. Backbone Northern European Model

Backbone is an open-source multi-energy and multi-sector optimisation framework built in GAMS[1]. Geographical and time resolution are user defined and set independently from the model framework. The model optimises the operation of the energy system and can be used in investment mode, where it can invest, e.g., in line-capacity reinforcement and installed capacity. Backbone is a least-cost optimisation model framework, but has also been used in a multi-objective setup. The model has been validated against other commercial and non-commercial models [64] used in international studies presented to the G7 [65], and in numerous use cases for hydrogen [66], hydrology and hydropower [67], power-to-X technologies in the energy system [66].

Backbone-NEM is a model dataset that can be run in Backbone and includes definitions of 16 European countries, extending from a Nordic perspective to a broader view that includes the European power market. It divides the regions by bidding zones as defined in ENTSO-E, and multiple scenarios are defined under it, namely the decentralised and the national trends scenario. Both are based on the TYNDP scenario and set the exogenous data for installed capacity. The model then optimises hourly production to minimise system costs. It includes a set of centralised production technologies, municipal district heating systems, and technology-specific exogenous data. In the current setup, the NEM omits electrolysers, as hydrogen demand has not been defined for the near-future scenario. The model includes limited carbon capture and storage (CCS) technologies, mainly in the UK, and these are therefore outside the scope of this model inter-comparison. It includes the weather data from 2009, as applied in the TYNDP national trends scenario.

[1] https://gitlab.vtt.fi/backbone/backbone

#### 2.1.2. Balmorel

Balmorel is a sector-coupled partial equilibrium energy system model that optimises investment and dispatch decisions across the European energy. Balmorel has been developed since 2001 and has increased in complexity (sectors, technologies, and spatio-temporal resolution) to address new challenges in the energy system. Balmorel has been compared and verified through model- or scenario- comparison studies [15, 16, 68]. The model includes production, energy infrastructure such as transmission and storage, and energy demands for the included sectors. Balmorel includes exogenously defined capacities for existing energy technologies and infrastructures and can perform transition pathways studies where the capacity development is optimised. Balmorel has widely been used in scientific articles and has been applied for pan-European studies, regional and national studies around the world, and down to local level. Several Balmorel model versions exist, which in this study are exemplified by Balmorel-DTU and Balmorel-NMBU, which have been used by activating different features and using different scenario assumptions:

- Balmorel-DTU: The model version used in this study is based on Kountouris et al. [69], which is the sector-coupled electricity, heating (district heating, individual heating), industry and hydrogen (representing high-value chemicals such as fuels and other). In this study, the geographical scope is limited to the Nordic countries. The decarbonisation pathway is facilitated using an increasing carbon price towards 2050.
- Balmorel-NMBU: In this study, the model represents the integrated Northern European power and district heating sectors. The transition to a renewable energy system is achieved through rapidly increasing carbon prices, rising to infinity in 2050. The model does not include negative emissions or CCS technologies; therefore, it heavily invests in renewable and storage technologies to meet demand. Furthermore, the energy demand is exogenously defined, but we allow some flexibility in the timing of demand through demand response [70] and smart charging [71]. Weather data from 2013 is used to ensure a realistic correlation between renewable resources and demand.

#### 2.1.3. highRES

highRES is a high-resolution electricity system optimisation model that integrates investment and dispatch decisions to identify cost-effective future electricity systems which usually show high levels of variable renewables. It uses hourly capacity factor time series aggregated from grid-cell level (30 km ERA5 resolution) to NUTS2, with demand-supply balancing and transmission modelled at the national level. Applied to the EU-25 (excluding Malta and Cyprus), Norway, Switzerland, and the UK, highRES also accounts for technical, environmental, and social land-use constraints on new wind power and solar photovoltaics (PV) projects. Results used in this paper are based on a medium scenario (in terms of socio-environmental land-use exclusions) from the WIMBY project [72].

In this study, the model includes existing electricity infrastructure expected to remain online in 2050 (e.g., the Olkiluoto nuclear plant in Finland). Weather data input used to generate capacity factor time series for solar, wind, and hydropower is based on ERA5 and the historical weather year 1995, which represents an average case in a European context [73]. A strict net-zero scenario is also implemented, based on a soft-coupling with a whole-energy system model (JRC-EU-TIMES), which represents the electricity sector's contribution to a net-zero energy system. highRES provides land availability scenarios to JRC-EU-TIMES, which then optimises the net-zero transition pathway to 2050 and returns the boundary conditions (electricity demand, CO2 emission targets) for use in highRES. highRES then optimises the spatially explicit investments and hourly operations of the European electricity system. Most countries have negative emission budgets, but a few have zero or positive, leading to a system-wide negative $CO_2$ budget of approximately 80 Mt across the modelled countries [72]. Negative emission technologies include both combined-cycle gas turbines with CCS and bioenergy with carbon capture and storage (BECCS), with the broader scenario design, model-coupling procedure, and implementation assumptions described in [72].

#### 2.1.4. IFE-TIMES-Norway

IFE-TIMES-Norway [59] is a long-term optimisation model which minimises total system costs and represents the full Norwegian energy system considering detailed techno-economic characterisation of resources, conversion, end-use technologies and energy service demands across the industry, buildings, and transport sectors. The model aggregates the Norwegian energy system into five spot market areas, oil and gas regions, and offshore wind regions, and models milestone years between 2018 and 2060. Cross-border electricity trade is characterised using exogenous electricity prices in countries outside Norway.

#### 2.1.5. LUT-ESTM

The LUT Energy System Transition Model (LUT-ESTM) [62, 74] describes a full energy-industry system with the ability for carbon dioxide removal [75]. All kinds of transition pathways can be described by scenarios, whereas the tool is designed to represent a high degree of overall electrification (direct, indirect) due to power-to-X routes [43], while the overall energy services demand is further growing. LUT-ESTM is applied in hourly resolution, multi-node, optimisation, with a variety of constraints, and based on a reproduced legacy system of the present. In total, about 170 technologies and variations are implemented. Solar PV prosumers are considered in combining a micro-economic perspective. Scenarios can assume that Europe aims for energy sovereignty, which may also lead to net exports from the Nordic countries. All Nordic countries are covered by LUT-ESTM [62, 74]. LUT-ESTM has been used for more than 75 articles from global down to local (city, island) level. LUT-ESTM has been identified as one of the most used tools for highly renewable energy system analyses [76] and suitable for very high standards in energy system transition research [24].

#### 2.1.6. ON-TIMES

ON-TIMES is an energy system optimisation model based on the TIMES framework. The model represents the Nordic region and was used for the Nordic Clean Energy Scenarios developed for Nordic Energy Research in 2021 [52, 77]. The model includes investments and operations of technologies used to serve the energy service demands of industry, buildings, and transport, as well as the supply sector (including renewable fuels) and electricity and district heat generation. The model includes Denmark, Norway, Sweden, Finland, and Iceland. The system is allowed to export and import fuels to the EU, and existing electricity transmission lines are modelled between the countries inside and outside the model scope. The scenario (Carbon Neutral Nordic) used in this paper is set up to run towards a 2050 goal of carbon neutrality, considering national plans, strategies, and targets.

### 2.1.7. OSeMBE

OSeMBE (Open Source electricity Model Base for Europe) [60] is an electricity system optimisation model developed using OSeMOSYS [78], a deterministic, least-cost optimisation tool designed for modelling long-term energy investment pathways. The model encompasses all 27 EU member states, as well as Norway, Switzerland, and the UK. The modelling period covers the 2015-2050 period and features a temporal resolution consisting of 15 distinct time slices.

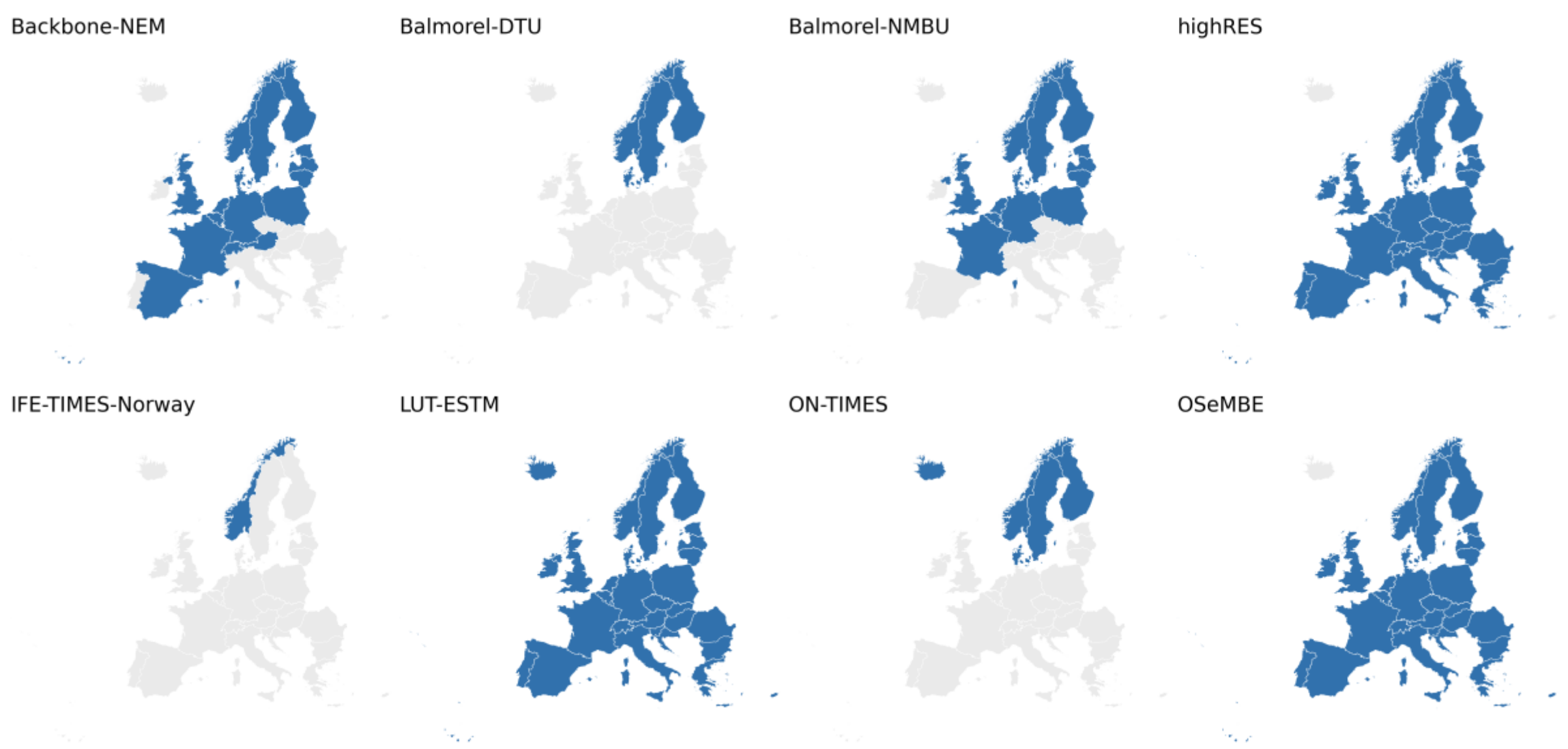

*Figure 2. Regional coverage of each model as used in this study. Note that some models have a broader geographical scope in their full configuration than what is shown here.*

## 2.2. Selection of Scenarios

Each modelling group selected existing scenarios that meet the criteria of 'climate neutrality' by 2050 or earlier. Modelling groups did not create new scenarios to participate in the model-comparison exercise, as the aim is to compare existing scenarios from recent or ongoing research projects. Since most teams have multiple scenarios that meet these criteria, the selection of the scenario was refined by examining the narrative aspect. For example, neutrality scenarios driven by technology change and adoption rather than policy-driven or societal commitments.

The scenarios examined in this study describe a broad set of possible future pathways. They vary in how they define the regional scope, the level of technology detail, and the core assumptions about system development. These differences reflect a range of decarbonisation approaches and spatial perspectives, forming a shared foundation for comparing how the participating models represent energy system transitions.

Models differ in how scenarios and assumptions are selected, reflecting a balance between normative and descriptive approaches. In this context, a normative approach typically allows the model to identify techno-economically optimal outcomes, whereas a descriptive approach incorporates stated or expected future policies in the regions modelled. Similar choices arise for other technologies, such as assumptions on expansion limits or policy targets, and their implementation often reflects the modeller's understanding of local conditions. These constraints are not always binding, as the model may avoid a technology even if it is available. When they are binding, however, they can help explain differences in results across models.

Backbone-NEM applies decentralised and national trends scenarios inspired by the TYNDP scenario, including exogenous capacity data and hourly optimisation of system costs, while omitting electrolysers due to the absence of hydrogen demand in the near-future scenario. Just like Backbone-NEM, the Balmorel-DTU and Balmorel-NMBU models both employ carbon price–driven decarbonisation pathways. Balmorel-DTU applies a gradually increasing price towards 2050, while Balmorel-NMBU enforces a rapid increase to infinity by 2050, resulting in high investment in renewables and storage in the absence of CCS technologies.

Explorations of net-zero pathways in the highRES model include soft-coupling with JRC-EU-TIMES, combining investment and dispatch optimisation under land-use, technical, and social constraints. It assumes a system-wide negative $CO_2$ budget of around 80 Mt across the modelled countries. The TECH scenario, explored using the IFE-TIMES-Norway model, represents a socio-technical pathway characterised by rapid technological progress and large-scale deployment of low-carbon technologies. It assumes high technology learning rates and high deployment potentials for renewable energy technologies (including onshore and offshore wind and solar PV), hydrogen, energy storage, CCS, and ammonia use in shipping. The scenario further assumes medium to high growth in energy service demand, largely driven by expanding industrial activity, combined with medium carbon prices and limited availability of bioenergy resources, with correspondingly high import costs for these. In addition, the TECH scenario is characterised by ambitious levels of hydrogen deployment, including the production of blue hydrogen, i.e., hydrogen production with $CO_2$ capture and storage, to support large volumes of hydrogen exports to Europe [79].

The LUT-ESTM scenarios describe highly electrified transition pathways with growing energy service demand and power-to-X integration. This approach envisions Europe-wide energy sovereignty with potential net exports from the Nordic countries. The ON-TIMES Carbon Neutral Nordic scenario models a regional energy system consistent with national strategies and a carbon-neutral target by 2050. The scenario encompasses all major end-use sectors, applies an emissions limit as part of a net-zero implementation, and allows fuel trade with the EU.

Finally, in the OSeMBE model, the net-zero case is defined by a gradual increase in carbon prices towards 2050. Emission reductions are mainly achieved using CCS technologies together with a continued expansion of VRE capacity. Because the model does not include trade expansion, large-scale storage, or hydrogen-based flexibility, CCS (particularly through BECCS) becomes the main way to offset the remaining emissions across the aggregated EU27 + NO/CH/UK system.

### 2.3. Temporal and Regional Scope

The results comparison is conducted at an annual timescale for the "Nordic4" countries, excluding Iceland. The "Nordic4" countries include Denmark, Finland, Norway, and Sweden. Two models (ON-TIMES and LUT-ESTM) in the comparison also cover Iceland, but none includes the islands and territories of the Nordic countries (such as Svalbard and the Faroe Islands). Balmorel-NMBU, Balmorel-DTU, IFE-TIMES-Norway, ON-TIMES, and OSeMBE all include some degree of aggregated time-step representation. In contrast, Backbone-NEM, highRES, and LUT-ESTM all employ an hourly temporal resolution. The OSeMBE model features yearly reporting for the 2020-2050 period, while Backbone-NEM and highRES provide snapshots for the years 2040 and 2050, respectively. Balmorel DTU, IFE-TIMES-Norway, ON-TIMES, and LUT-ESTM report their results in 5-year intervals, and Balmorel-NMBU reports in 10-year increments.

The models' regional scope ranges from the national to the continental level, as shown in Figure 2. IFE-TIMES-Norway is the only national-level model, while in this study Balmorel-DTU and ON-TIMES cover the Nordics. Furthermore, Backbone-NEM and Balmorel-NMBU expand the Nordic coverage to include most Western European countries. The models with the broadest geographical scope include highRES, LUT-ESTM, and OSeMBE, including most of continental Europe.

### 2.4. Selected Result Variables

The modelling teams reported a set of variables from their climate-neutrality scenarios, with outputs recalculated to align with the IAMC template using *pyam*[2]. This step ensured consistent units and harmonisation across the models. Variables were selected from the IAMC variable list [80], based on the OpenENTRANCE nomenclature. The analysis focuses on a subset of key variables, including final electricity demand, installed capacities of major generation technologies, secondary electricity, carbon capture, and $CO_2$ emissions.

Because the comparison is non-harmonised, the common reporting structure should not be interpreted as a shared scenario protocol or common input dataset. Results are compared using common output indicators, units, countries, and reporting years where available, while input assumptions, model scope, technology representation, and scenario design vary across models. The main dimensions of comparability and non-harmonisation are summarised in Supplementary Table S2 and should be considered when interpreting differences across the model ensemble.

[2] Tutorial on unit conversion: https://pyam-iamc.readthedocs.io/en/stable/tutorials/unit_conversion.html

Electricity demand is shown separately in Figure 3 because it is an important contextual variable for interpreting capacity expansion, but it is not treated consistently across all models. In some models, electricity demand is an exogenous scenario assumption; in others, particularly full-energy-system models, it is an endogenous outcome of sectoral optimisation. Demand values should therefore be interpreted as part of each model-scenario context rather than as a harmonised input.

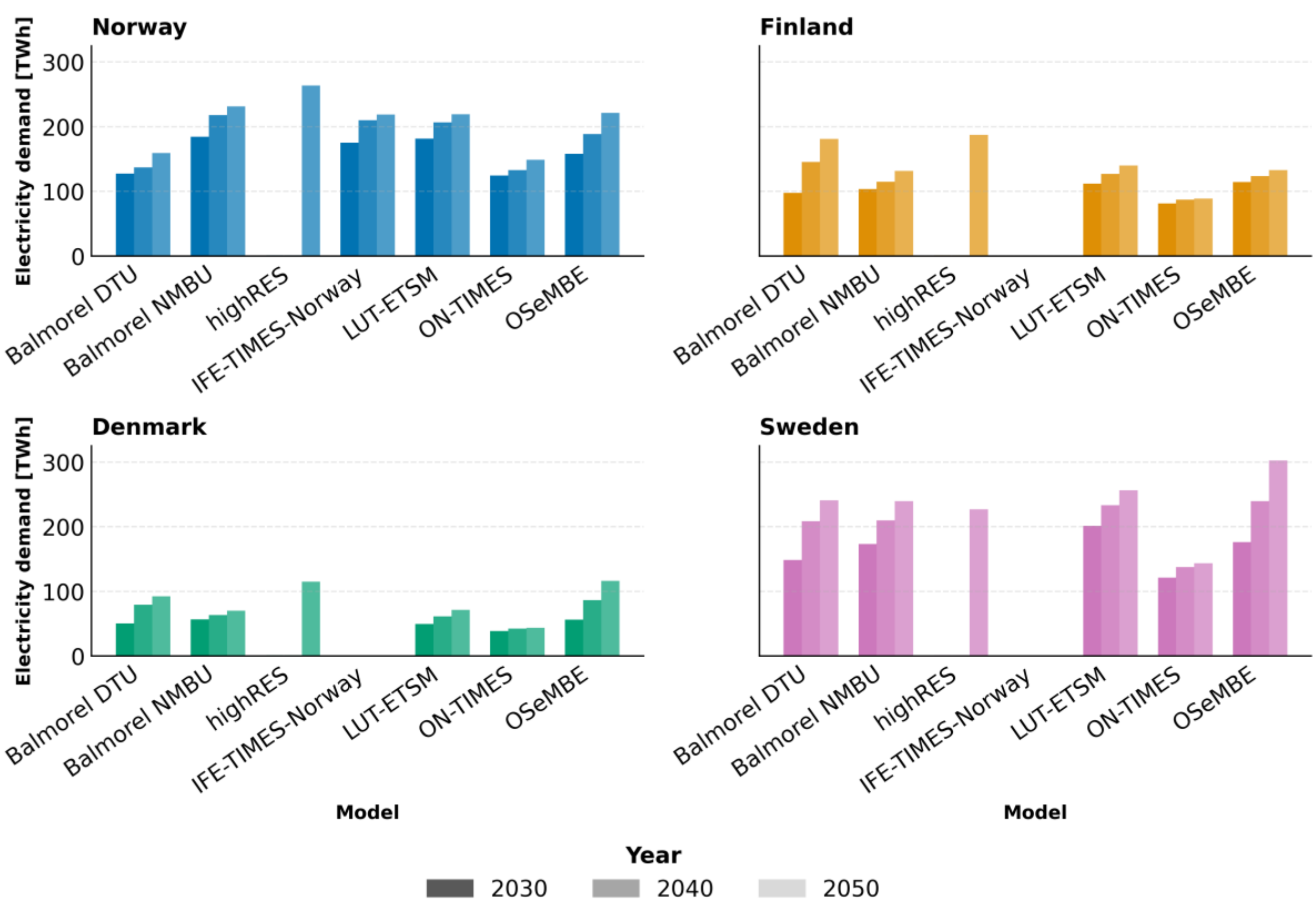

*Figure 3. Reported final electricity demand by country and model, 2030–2050. Demand is exogenous in some models and endogenous in others; values therefore represent model-scenario context rather than harmonised input assumptions.*

Figure 3 presents the reported electricity demand for Denmark, Finland, Norway, and Sweden. Backbone-NEM is excluded because it does not provide country-level values for this variable. The figure illustrates both the comparability enabled by common reporting and the limits of comparability in a non-harmonised ensemble. For example, Balmorel-NMBU, IFE-TIMES-Norway, and OSeMBE report broadly similar demand levels in several countries, while Balmorel-DTU reports higher values for Denmark. In IFE-TIMES-Norway and ON-TIMES, the demand shown is an output from full-energy-system optimisation rather than an exogenous input assumption.

## 3. Results and Discussion

This section presents the model results for selected power generation technologies, including CCS and the associated emission levels, for the years 2030, 2040, and 2050. The analysis highlights where the models produce similar outcomes and where they differ, and relates these patterns to the assumptions that shape each modelling framework. These points help frame the interpretation of the results and clarify the factors that drive variation across the model ensemble. Please note that not all models report values for all years. Each boxplot reflects the distribution across the models that report results for that decade.

It is important to note that, in a non-harmonised multi-model comparison, differences in results arise from a combination of structural modelling choices (e.g., temporal resolution, sectoral scope, optimisation framework) and input assumptions (e.g., technology costs, resource potentials, policy constraints). These effects are not isolated in the present analysis, but are instead considered jointly, reflecting how differences between modelling studies are encountered in practice.

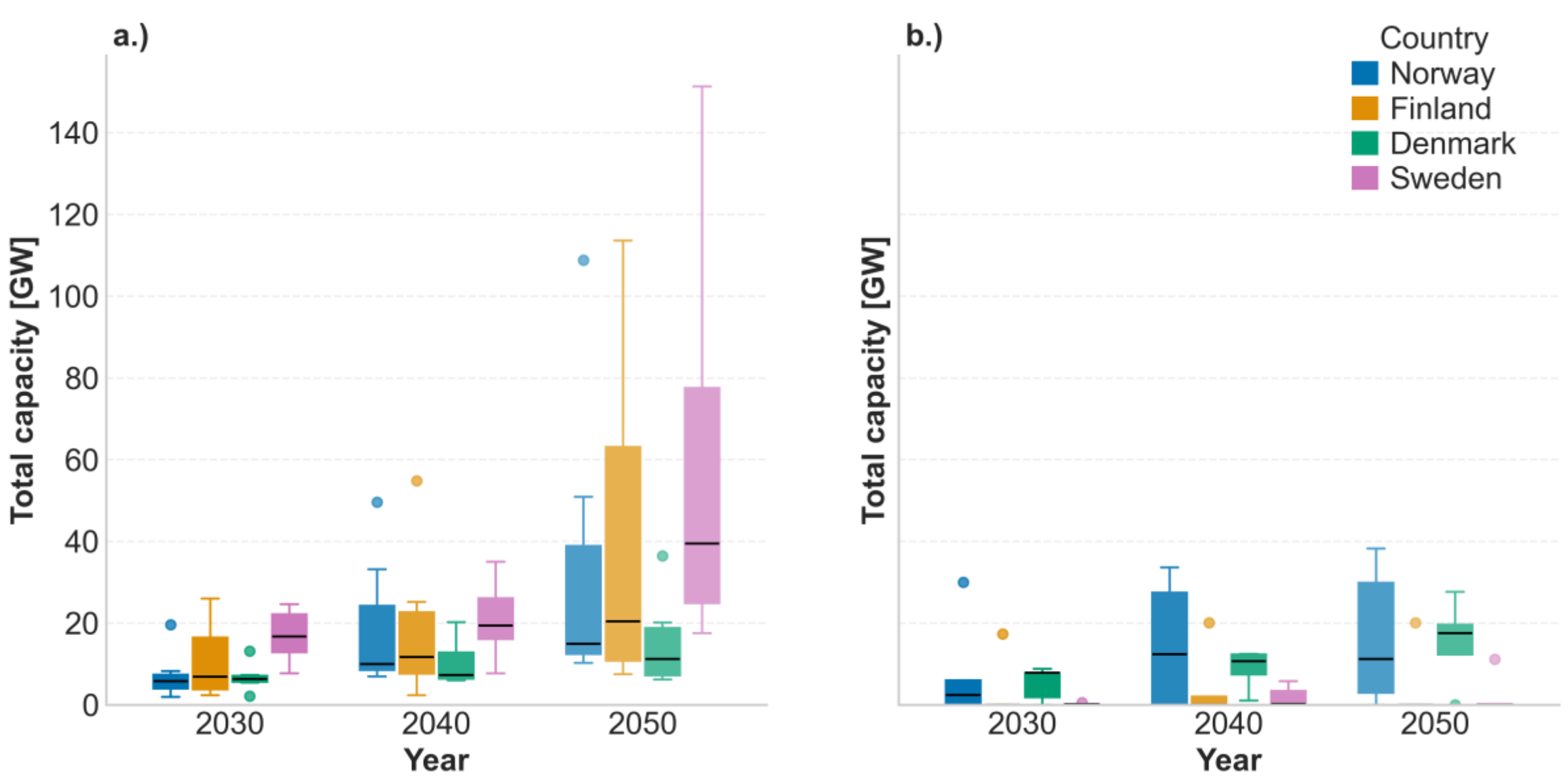

*Figure 4. Onshore (a) and offshore (b) wind power capacity across Nordic countries in 2030, 2040, and 2050. Boxplots display the distribution of model results for Norway, Finland, Denmark, and Sweden in 2030, 2040, and 2050; horizontal lines mark the median, boxes indicate the interquartile range (IQR), whiskers extend to 1.5 × IQR, and points represent outliers. Total capacity is expressed in gigawatts (GW). Each boxplot represents results from up to eight models, depending on data availability for each year and technology. Full details of the results are available on Zenodo [81].*

Across the models, both the assumed onshore wind potential and the resulting deployment shown in Figure 4a differ markedly. A subset of models, Balmorel-DTU, Balmorel-NMBU, LUT-ESTM, ON-TIMES, and OSeMBE, reach some of their stated potential limits, and even then, only in specific countries and years. Balmorel-NMBU exhausts its potential in most cases, with the exceptions being Finland and Sweden in 2030. LUT-ESTM maxes out its potential for all countries in 2050, while OSeMBE reaches its constraint in Finland in 2030 and in 2050. ON-TIMES reaches its potential limit for Denmark in 2040 and 2050. In general, the upper capacity potential assumptions applied in Balmorel-NMBU, Balmorel-DTU, ON-TIMES, and OSeMBE are lower than those used in highRES and LUT-ESTM, although OSeMBE shows higher values for Denmark in certain years. These differences become most apparent in the 2050 results, where highRES, LUT-ESTM, and OSeMBE consistently fall at the upper end of the range of modelled onshore wind deployment. This highlights the critical role of VRE potential estimates in modelling outputs.

Across scenarios, wind power emerges as a cornerstone of Nordic net-zero power systems. By 2050, onshore wind power provides about 40% of total capacity, with offshore wind power contributing a further 12%, such that wind power collectively represents more than half of mid-century capacity across models. These findings support recent examinations of pathways to a climate-neutral Europe by 2050, which suggest that models align on the shift toward renewables by mid-century [12]. Figure 4 shows the distribution of wind power capacity by country. Onshore wind power expansion is evident in all countries between 2030 and 2050. Substantial offshore deployment appears only in Denmark, where the interquartile range narrows over time, indicating growing agreement across models. In Norway, offshore capacity also increases, but the interquartile range widens towards 2050 (Figure 4b), reflecting greater divergence in model outcomes and less robust evidence than for Denmark. Unlike other countries, due to Norway's topology it requires floating offshore technology already close to shore. The results indicate that different models have different economic considerations regarding the competitiveness of floating offshore. For example, IFE-TIMES-Norway allows offshore wind investments to electrify offshore oil and gas infrastructure, and offshore wind can be used to supply electricity domestically and/or be exported to the European power market, while onshore wind potentials remain limited. Given the scenario assumptions, including high activity levels in end-use sectors, IFE-TIMES-Norway has the highest offshore wind capacity in Norway among all models. In contrast, wind power expansion in Finland and Sweden remains predominantly onshore. By 2050, only three models report any offshore wind capacity in Sweden, and only two in Finland.

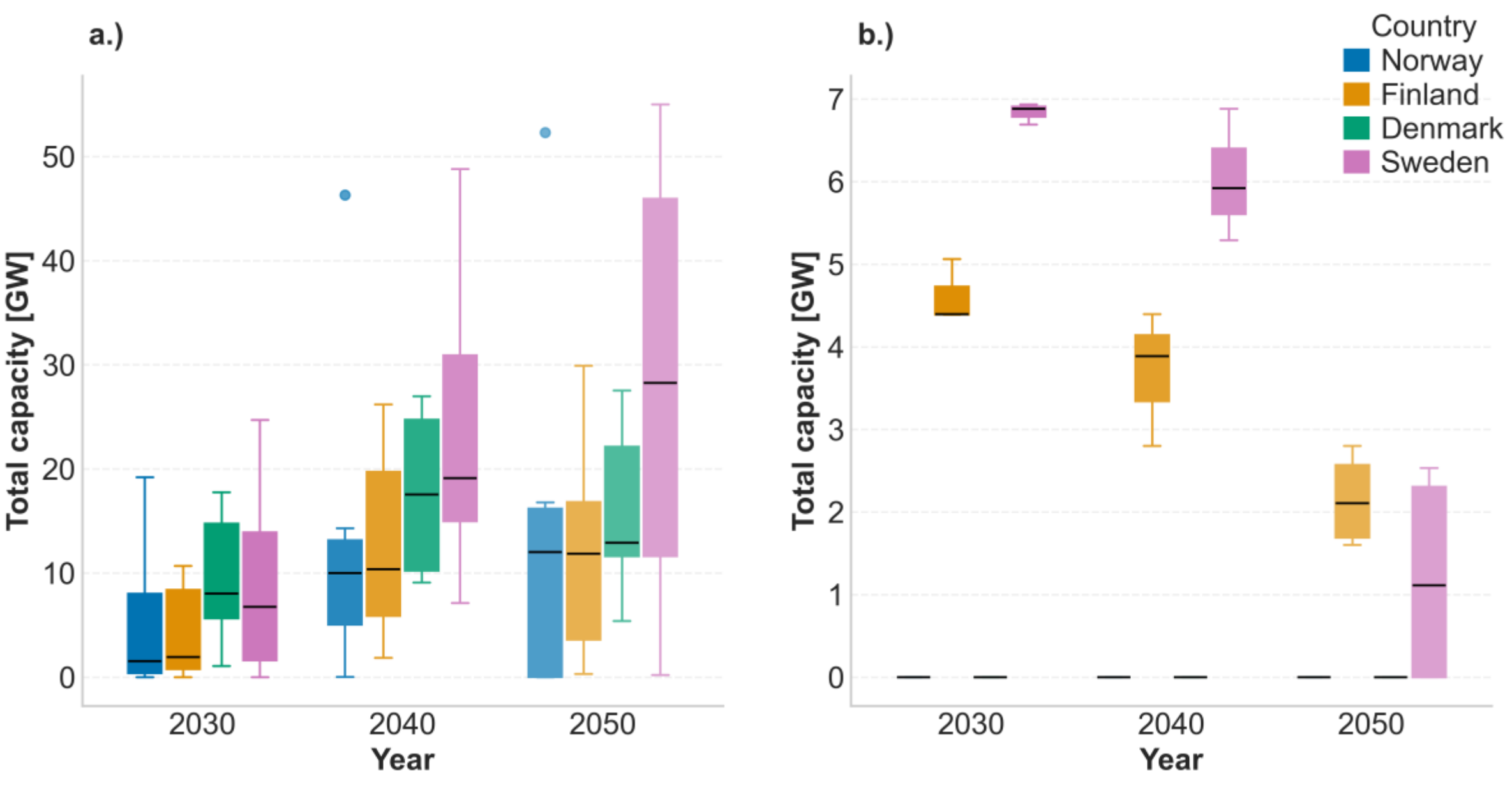

*Figure 5. Total capacity of solar PV (a) and nuclear power (b) by country and year.*

LUT-ESTM's behaviour in our ensemble reflects a few structural choices that shape its treatment of nuclear and renewable technologies. Although LUT-ESTM allows potential investment in nuclear power in some instances, even when permitted, the model does not invest in it because the technology is not competitive under current cost assumptions. This outcome is consistent with recent assessments that show the economic challenges of new nuclear [82]. In the scenarios used here, new nuclear construction was not allowed, although existing plants operate 40, and in some cases 60 years, and permitting new plants would not materially change the results for the same economic reason.

Similar conclusions have been reached in recent system-level assessments for the Nordic region. For Denmark, Thellufsen et al. [83] showed that a fully renewable, sector-coupled energy system is less costly than nuclear-based alternatives under current cost assumptions. Building on this work, Moen et al. [84] demonstrated that while nuclear power can become more cost-competitive under assumptions of higher renewable costs and reduced wind capacity factors, narrowing the cost difference relative to renewable-dominated systems, total system costs remain broadly similar across different nuclear shares, with only modest differences depending on assumptions regarding sector coupling and utilization. For Sweden, Kan et al. [85] similarly found limited economic rationale for reinvesting in nuclear power once existing reactors are decommissioned. Taken together, these studies suggest that new nuclear investments are unlikely to play a major role in cost-optimal Nordic energy systems under current techno-economic assumptions.

Across the model ensemble, nuclear power declines in the Nordic region by 2050 (Figure 5b), with decreasing capacity in Finland and Sweden and no deployment in Denmark or Norway. This outcome reflects generally limited nuclear investment across the selected scenarios, together with differences in how existing capacities, plant lifetimes, and new investment options are represented. Several models restrict new nuclear construction based on national policy assumptions, while others technically permit nuclear expansion in some countries but do not select new capacity under the scenario assumptions applied here. As a result, no model deploys new nuclear additions during the modelling horizon.

By 2050, residual nuclear capacity remains only in a subset of models and countries. Finland retains capacity in OSeMBE, LUT-ESTM, Backbone-NEM, and highRES, while Sweden retains capacity only in OSeMBE. Cross-model differences therefore mainly reflect assumptions on existing capacity retention, plant lifetimes, policy constraints, and technology representation, rather than a systematic assessment of new nuclear competitiveness. Because Backbone-NEM, highRES, IFE-TIMES-Norway, and OSeMBE do not include small modular reactors, and most models represent only conventional nuclear technologies, these results should not be interpreted as general conclusions about all possible nuclear futures. Model-level assumptions on nuclear and hydropower capacity investment are summarised in Supplementary Figure S1.

Renewable expansion in the LUT-ESTM is governed by a limit on the annual change in renewable capacity share, capped at four percentage points per year. This approach, based on observed deployment trends, avoids unrealistic jumps in deployment while still allowing rapid growth. The model also includes an explicit representation of solar PV prosumers, differentiating between residential, commercial, and industrial prosumers, optimised with their own battery systems. As prosumers are often omitted or simplified in other large-scale models, this added granularity can contribute to differences in solar outcomes relative to the rest of the ensemble.

In highRES, solar PV deployment is constrained by an annual buildout rate and a land availability restriction. However, the modelled land availability is far larger than the rate-based cap, meaning that the latter limits PV deployment. This assumption is grounded in historical expansion patterns and effectively restricts the pace at which solar capacity can grow. In Balmorel-DTU and Balmorel-NMBU, potentials for VRE sources are represented by three so-called resource grades, which represent the best, second and third locations and priorities. ON-TIMES model uses cost assumptions for solar PV that reflect 2021 values, which could be higher than those applied in other models. This modelling choice may contribute to the relatively limited deployment of solar PV observed in the ON-TIMES results. Except for the ON-TIMES projections for Norway in 2040 and 2050, solar PV deployment in the remaining models is not constrained by assumed resource potentials. Consequently, the wide range of solar PV capacities shown in Figure 5a is primarily attributable to differences in techno-economic assumptions—such as investment costs and capacity factors—as well as to broader variations in system representation and model structure.

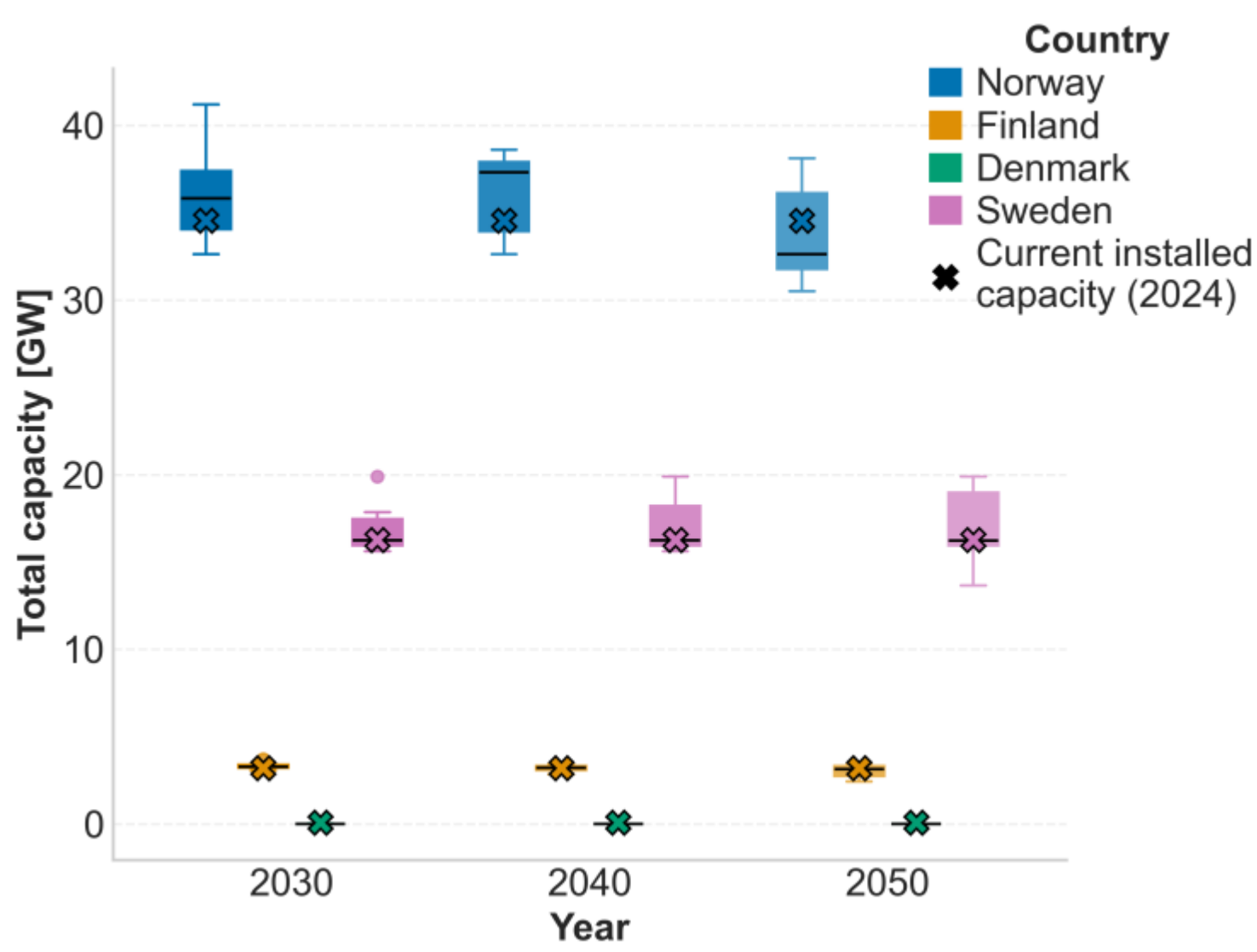

*Figure 6. Total capacity of hydropower by country and year.*

Hydropower remains a cornerstone of the Nordic power system, particularly in Norway and Sweden, where installed capacity is much higher than in Denmark and Finland. Across the ensemble, hydropower capacity changes only marginally between 2030 and 2050, as shown in Figure 6. This mainly reflects assumptions that most existing plants will remain in operation and that large-scale decommissioning is unlikely. Although hydropower facilities face ageing-related refurbishment needs, historical patterns show that major plants are typically upgraded or relicensed rather than permanently retired. This results in stable or occasionally increased capacity over time.

In practice, future hydropower capacity could exceed what the models show. Uprating through dam heightening, new turbines, or smaller new installations could all contribute to this. These developments would not be cost-free, since refurbishment work, environmental measures, and licensing processes add to the investment burden. Even so, hydropower remains an essential source of system flexibility and balancing throughout the transition towards net zero, even if its share of total capacity gradually declines.

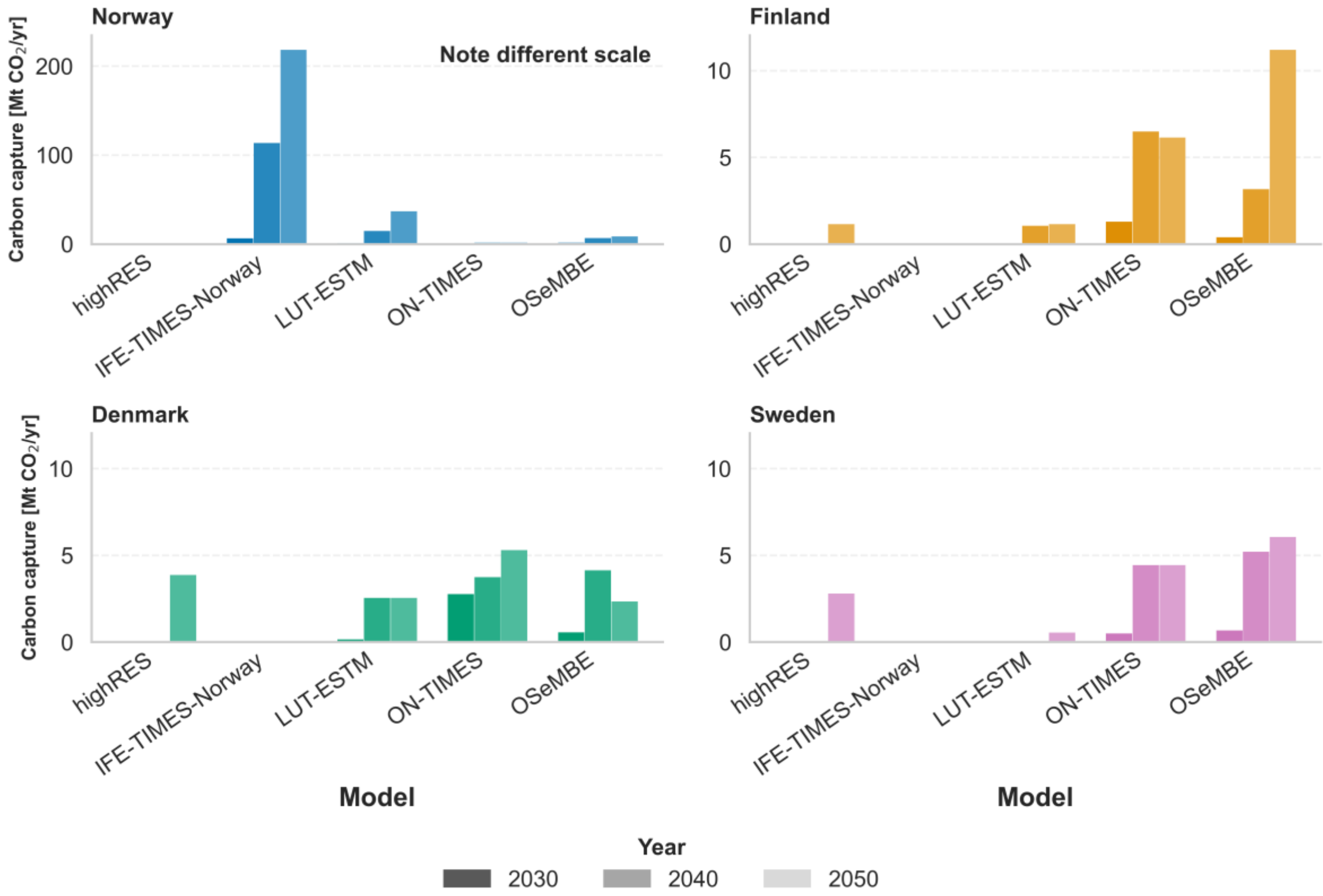

*Figure 7. Total carbon capture and sequestration expressed in Mt $CO_2$ per year, across models and years. Note, the y-axis for Norway is significantly higher compared to the other countries due to large-scale blue hydrogen production for the selected scenario.*

The IFE-TIMES-Norway, ON-TIMES, and OSeMBE models report CCS deployment at the country level. By 2030, ON-TIMES and OSeMBE show broadly consistent levels of CCS uptake, with relatively narrow ranges across countries. Beyond 2030, however, the spread in results increases markedly, particularly by 2050, reflecting less convergence on the long-term role of CCS in Nordic energy systems. A notable feature is the IFE-TIMES-Norway model, which reports substantially higher CCS deployment related to highly ambitious levels of blue hydrogen (produced using natural gas and coupled with CCS) production for exports after 2030 compared to the other models, driving the numbers up for the case of Norway, as shown in Figure 7. The results shown in Figure 8 highlight the sensitivity of CCS deployment outcomes to modelling assumptions, such as the inclusion of blue hydrogen production.

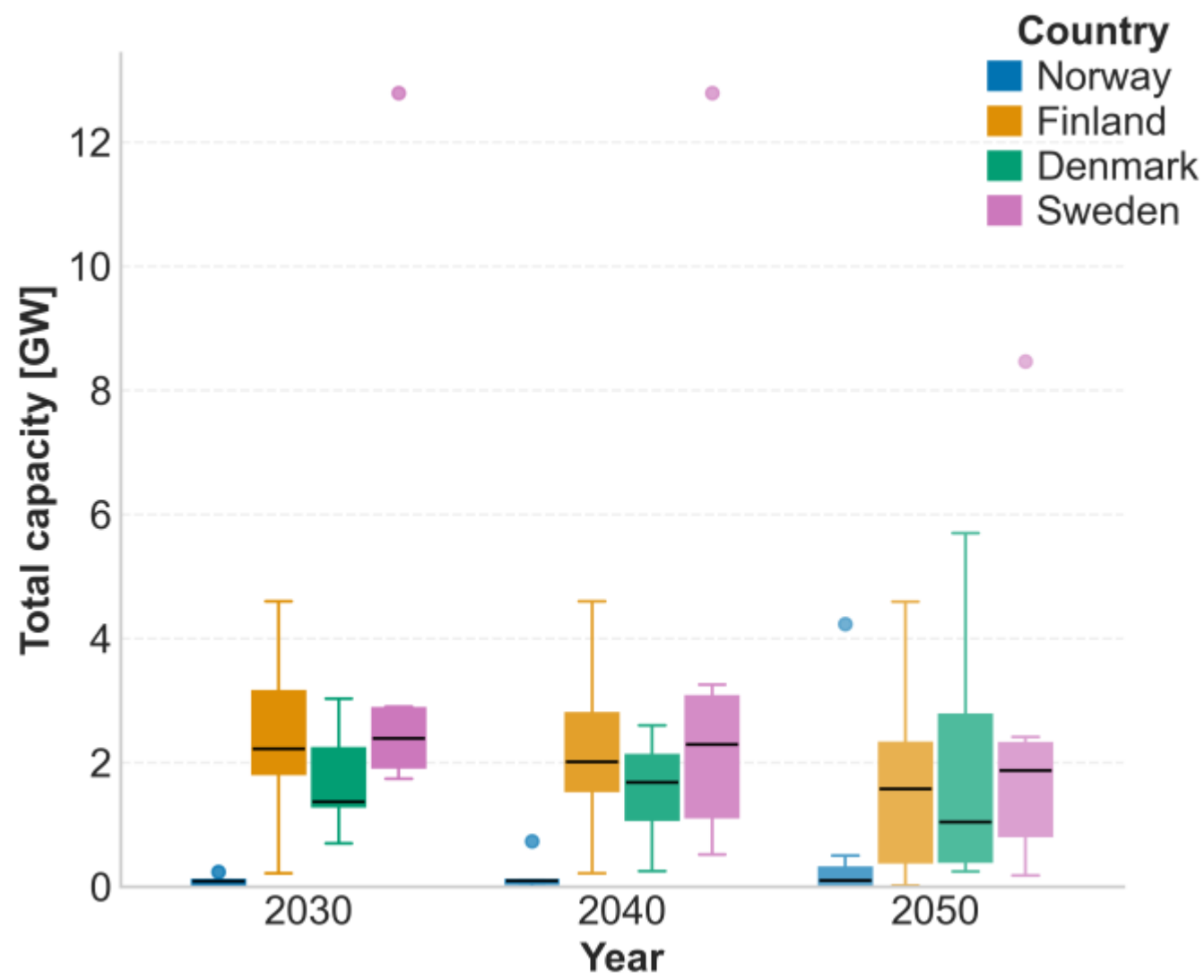

*Figure 8. Total capacity of biomass power plants by country and year.*

Model results indicate that biomass-fired power capacity is concentrated mainly in Sweden, Finland, and Denmark, with Sweden showing the highest individual model outcomes but not consistently dominating across the ensemble. By 2030, Sweden and Finland are generally at comparable levels across reporting models, while Denmark also contributes substantial capacity. Norway exhibits the largest relative increase in biomass capacity, albeit from a low base. For Denmark and Sweden, most models (Balmorel-NMBU, Balmorel DTU, and ON-TIMES) show declines from 2030 to 2050, while OSeMBE stands apart with higher values in 2050, visible as outliers in Figure 8. Overall, the multi-model evidence points to contraction in Denmark and Sweden, with OSeMBE providing a contrasting scenario.

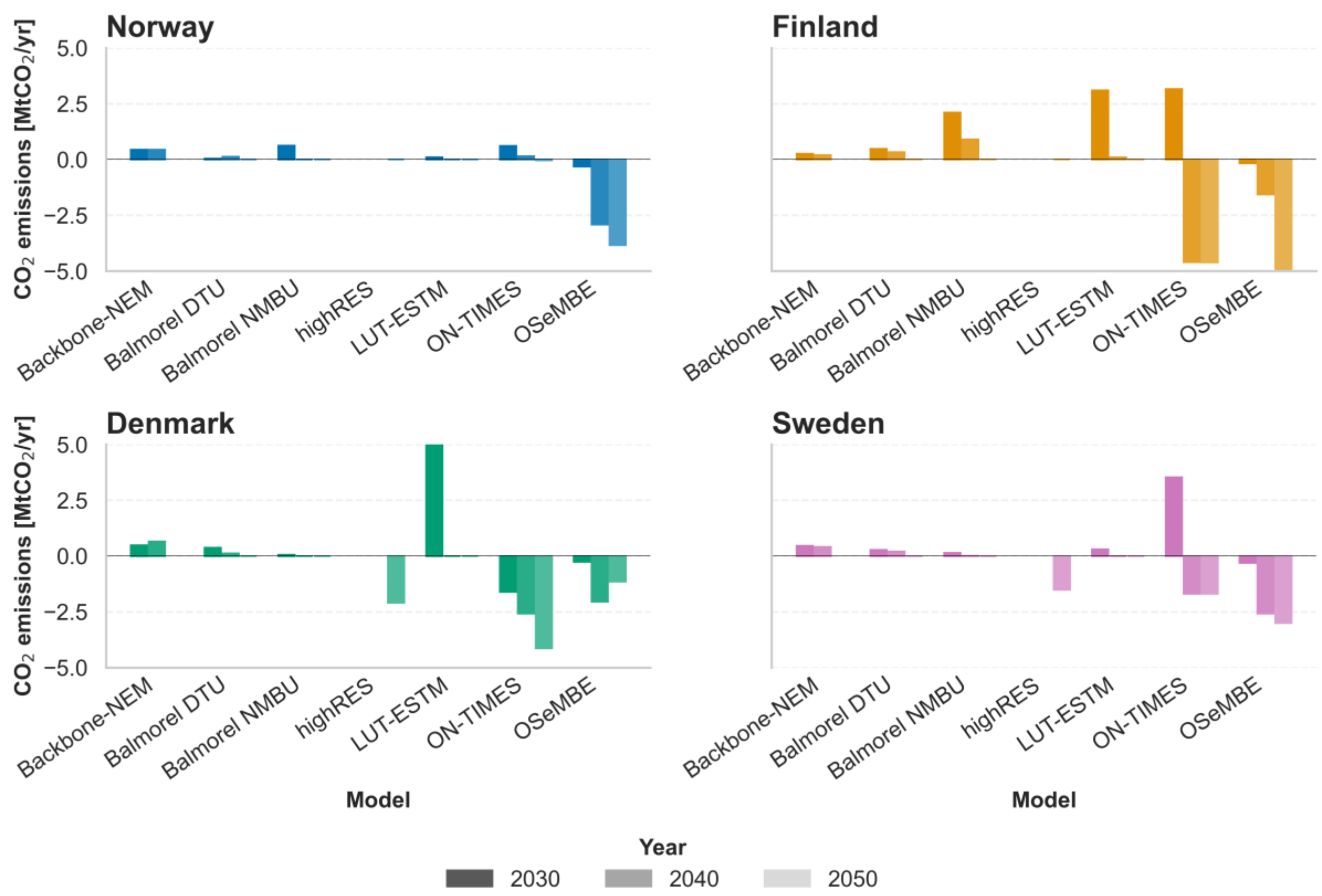

*Figure 9. Power-supply $CO_2$ emissions by country, model and year, given in $MtCO_2$/yr.*

Differences in how models treat a 2050 net-zero framing lead to divergent interpretations of 'net zero'. As shown in Figure 9, some models report residual emissions in 2050, others reach exact neutrality, and a few exhibit net negative emissions. The latter outcome arises in OSeMBE through the deployment of biomass-fired power plants with CCS. While the magnitude of this deployment is considerable relative to other models, it reflects the system-wide optimisation of the pan-European OSeMBE framework, which allocates substantial biomass use to the Nordic region. In contrast, other regions in the model employ little or none.

The prominent role of biomass with CCS in OSeMBE stems from several interacting model features. Among the available options, biomass combined with carbon capture offers the greatest potential to reduce emissions. Under the high $CO_2$ prices projected toward the end of the modelling horizon, the system's cost minimisation tends to favour its deployment. Additionally, the technology helps meet the reserve-margin requirement, further enhancing its value within the modelled power system. Its extensive use can also be understood as a response to the simplified representation of other system elements, as OSeMBE includes neither large-scale energy storage nor hydrogen infrastructure, limiting the range of low-carbon flexibility options. Finally, the model accounts only for emissions from the power sector, making biomass with CCS an especially appealing option for achieving significant emissions cuts within that limited scope.

## 3.1. Policy Implications

The results highlight both the usefulness and the limitations of non-harmonised multi-model comparisons for policy. The main policy value of this study does not lie in identifying a single preferred pathway or precise capacity forecast, but in distinguishing between robust cross-model messages and outcomes that remain sensitive to model scope, assumptions, and technology representation.

The most robust policy-relevant finding is the central role of wind power in Nordic power-sector transitions. Across the model ensemble, wind power, mainly onshore but complemented by offshore wind in some countries, emerges as the backbone of the future power system. The policy implication is therefore not simply to target additional wind capacity, but to address the enabling conditions that determine whether large-scale wind deployment can be realised. Timely permitting, grid reinforcement, offshore spatial planning, and flexibility options become central planning priorities across a wide range of model outcomes.

At the same time, the large spread in installed capacities, CCS deployment, and emissions outcomes indicates that Nordic energy planning should not rely on single-model point estimates. Instead, model results are better used as ranges of plausible outcomes, that can support stress-testing and adaptive planning. Instead, model results are better used as ranges of plausible outcomes, particularly for decisions related to renewable buildout, transmission needs, offshore wind development, system flexibility, and technologies whose role varies substantially across models.

The comparison also underlines the importance of transparency in policy-facing modelling. Apparent disagreement between model results may reflect substantive differences in transition pathways, but it may also arise from differences in geographical scope, sectoral boundaries, demand treatment, technology availability, CCS and BECCS representation, and emissions-accounting boundaries. For policy use, model outputs should therefore be accompanied by contextual information describing these assumptions, so that differences in results can be interpreted correctly.

A further implication concerns the interpretation of “net zero”. The model results show that climate-neutrality framings are not operationally equivalent across models. Some scenarios report residual positive emissions, others reach exact neutrality, and others produce net-negative emissions through negative-emission technologies. For policy, this means that net-zero scenario results should specify whether neutrality is achieved within the power sector or across the wider energy system, whether negative emissions are allowed, and whether offsets occur inside or outside the reported country or sector boundary.

Finally, the results suggest caution when interpreting technologies whose deployment varies strongly across models, particularly CCS, BECCS, and nuclear power. The selected scenarios do not provide a basis for general conclusions about the future viability of all CCS pathways or emerging nuclear options such as SMRs. Instead, they show that these outcomes depend strongly on technology availability, policy constraints, sectoral scope, and emissions-accounting boundaries. When such technologies are central to a policy question, dedicated scenario or sensitivity analysis is required before drawing specific technology conclusions.

# 4. Limitations and Future Work

This study has several limitations that should be kept in mind when interpreting the results. First, the non-harmonised design means that each model retains its own assumptions, including differences in techno-economic data, policy constraints, and scenario narratives. This reflects how modelling studies are typically developed in practice, but it also means that differences in results cannot be attributed to single factors in a clear way. Instead, the outcomes reflect a combination of interacting modelling choices.

The models also differ in scope, geographical coverage, temporal resolution, and system boundaries. Some represent the full energy system, while others focus mainly on electricity. There are also differences in geographical coverage and temporal resolution. These aspects influence how flexibility options, trade, and technologies such as CCS and hydrogen are represented, and therefore contribute to variation in the results. In addition, not all models report the same variables, countries, technologies, or time steps, which limits the comparability of some indicators and means that the number of models represented differs across figures and years.

Finally, the study is based on existing scenarios provided by the modelling teams. No additional harmonisation or coordinated sensitivity analysis was carried out. This was consistent with the aim of comparing independently developed scenarios under native model assumptions, but it limits the ability to isolate the effect of specific assumptions or model features.

Future work could complement this approach with more controlled comparison designs. For example, targeted harmonisation of selected assumptions or coordinated sensitivity analysis could help identify which factors most strongly affect model divergence. Such work would answer a different question from the present study: not what range of outcomes emerges from independently developed Nordic transition scenarios, but how specific assumptions or model features shape that range. Combining these approaches would preserve the practical value of non-harmonised comparison while improving the ability to explain why model outcomes differ.

# 5. Conclusions

This study compared Nordic power-sector transition scenarios across eight structurally diverse modelling frameworks operating without harmonised inputs. Focusing on Denmark, Finland, Norway, and Sweden in 2030, 2040, and 2050, the analysis examined convergence and divergence in generation capacity, $CO_2$ emissions, and carbon capture and storage (CCS). To our knowledge, this is the broadest non-harmonised multi-model comparison of Nordic power-sector transition pathways to date in terms of the number of models and country-level indicators considered.

Across the ensemble, the clearest shared outcome is the substantial expansion of variable renewable energy. Wind power, predominantly onshore, accounts for the largest share of modelled capacity by 2050 in most scenarios, complemented by growing solar PV contributions. However, total installed capacities differ markedly across models, with differences interpreted considering renewable-resource potentials, techno-economic assumptions, deployment constraints, and model structure.

Hydropower capacity remains broadly stable across the modelling horizon, reflecting assumptions that most existing plants remain in operation. Nuclear power declines across the selected scenarios, with no model deploying new nuclear capacity. Differences in nuclear outcomes mainly reflect assumptions on existing plant lifetimes, policy constraints, and technology representation, and should therefore not be interpreted as general conclusions about all possible nuclear futures.

The strongest divergence appears for technologies associated with CCS and negative emissions. Biomass-fired power generation plays a limited role in most models, but OSeMBE shows substantial biomass with CCS deployment, leading to net-negative emissions. IFE-TIMES-Norway reports substantially higher CCS deployment in Norway, associated with large-scale blue-hydrogen production coupled with CCS. These outcomes illustrate how CCS results depend on sectoral scope, technology availability, flexibility representation, emissions boundaries, and scenario framing.

Emissions outcomes also show that "net zero" is not operationally equivalent across models. By 2050, results range from residual emissions to net-negative values, reflecting both technology choices and differences in emissions-accounting boundaries. Overall, the findings show broad agreement on the direction of Nordic power-sector transitions, alongside substantial variation in technology composition, capacity levels, CCS deployment, and emissions outcomes. The study therefore provides a synthesis of prevailing Nordic energy-modelling perspectives rather than a harmonised benchmarking exercise, and highlights the need to interpret non-harmonised model results in light of model scope, assumptions, and scenario design.

## CRediT authorship contribution statement

**Emir Fejzic**: Conceptualisation, Data curation, Formal analysis, Investigation, Methodology, Visualisation, Writing – original draft. **Will Usher**: Conceptualisation, Investigation, Methodology, Writing – review & editing, Funding acquisition. **Pernille Seljom**: Resources, Writing – review & editing. **Miguel Chang**: Resources, Writing – original draft, Writing – review & editing. **Oskar Vågerö**: Resources, Investigation, Writing - review & editing. **Maximilian Roithner**: Resources, Investigation, Writing - review & editing. **Marianne Zeyringer**: Resources, Writing - review & editing. **Guillermo Valenzuela**: Resources, Investigation. **Ida Græsted Jensen:** Resources, Writing – review & editing. **Eirik Ogner Jåstad**: Resources, Writing – review & editing. **Rasmus Bramstoft:** Resources, Investigation, Writing - review & editing. **Marie Münster**: Resources, Writing - review & editing. **Jean-Nicolas Louis**: Resources, Writing - review & editing. **Dmitrii Bogdanov:** Resources. **Christian Breyer:** Resources, Writing – review & editing.

## Funding

This work was carried out within the Nordic Climate and Energy Modelling Forum (NCEMF) and received seed funding from the KTH Energy Platform in April 2022. The development and application of the IFE-TIMES-Norway model were further supported by the FME NTRANS – Norwegian Centre for Energy Transition Strategies, grant number 296205, funded by the Research Council of Norway. The development of the highRES modelling results received funding from the European Union's Horizon Europe research and innovation programme under grant agreement no. 101083460 (WIMBY). DB and CB received funding from the CETP RESILIENT project under grant 2909/31/2023, and NER NORDH2ub project under grant 2074/31/2023.

## Declaration of competing interest

The authors declare that they have no known competing financial interests or personal relationships that could have appeared to influence the work reported in this paper.

## Data availability

Data supporting the results presented in this study are available at Zenodo, 10.5281/zenodo.18189762.

## Acknowledgements

The authors would like to thank Afzal Siddiqui (Stockholm University and Aalto University) for commenting on a draft version of this paper.

## Declaration of generative AI and AI-assisted technologies in the writing process

The authors used ChatGPT-5.5 (OpenAI) to improve the clarity, conciseness, and readability of the manuscript. The authors reviewed and edited the content as needed.